\documentclass[pre,aps,twocolumn]{revtex4}
\usepackage{epsfig} 
\usepackage{bm} 
\newcommand{\B}[1]{{\bm{#1}}}
\newcommand{\C}[1]{{\mathcal{#1}}} 
\renewcommand{\it}[1]{\textit{#1}}
\newcommand{\Onecol} {\begin{widetext} \onecolumngrid} 
\newcommand{\Twocol} {\end{widetext} \twocolumngrid}   
\begin{document}
\title{Drag Reduction by Polymers in Turbulent Channel Flows:\\ Energy Redistribution Between Invariant
Empirical Modes.}
\author{Elisabetta De Angelis$^{1}$,  Carlo M. Casciola$^{1}$, Victor S. L'vov$^{2}$,
Renzo Piva$^{1}$ and Itamar Procaccia$^{2}$}
\affiliation{$^1$ Dip. Mecc. Aeron., Universit\`a di Roma ``La Sapienza", 
Via Eudossiana 18, 00184, Roma, Italy\\
$^2$ Dept. of Chemical Physics, The Weizmann Institute of Science, Rehovot, 76100 Israel.}
\begin{abstract}
We address the phenomenon of drag reduction by dilute polymeric additive to turbulent 
flows, using Direct Numerical Simulations (DNS) of the FENE-P model of viscoelastic flows.
It had been amply demonstrated that these model equations reproduce the phenomenon,
but the results of DNS were not analyzed so far with the goal of interpreting the
phenomenon. In order to construct a useful framework for the understanding of drag
reduction we initiate in this paper an investigation of the most important modes that
are sustained in the viscoelastic and Newtonian turbulent flows respectively. The modes are
obtained empirically using the Karhunen-Lo\'eve decomposition, allowing us to compare
the most energetic modes in the viscoelastic and Newtonian flows. The main finding of
the present study is that {\em the spatial profile of the most energetic modes is hardly changed between
the two flows}. What changes is the energy associated with these modes, and their relative 
ordering in the decreasing order from the most energetic to the least. Modes that are
highly excited in one flow can be strongly suppressed in the other, and vice versa. This dramatic
energy redistribution is an important clue to the mechanism of drag reduction as is proposed
in this paper. In particular there is an enhancement of the energy containing modes in the
viscoelastic flow compared to the Newtonian one; drag reduction is seen in the energy containing
modes rather than the dissipative modes as proposed in some previous theories.
\end{abstract}
\maketitle
\section{Introduction}
``Drag reduction" refers to the interesting observation that the addition of a few tens 
of parts per million (by weight) of
long-chain polymers to turbulent fluids can bring
about a reduction of the friction drag by up to 80\% \cite{00SW}. Obviously, the
phenomenon has far reaching practical implications besides being challenging from
the fundamental point of view.
In spite of intense interest for an extended period of
time \cite{69Lu,75Virk,90Ge}, Sreenivasan and White \cite{00SW} 
recently concluded that ``it is fair to say that the
extensive - and continuing - activity has not produced a firm grasp of the mechanisms of drag reduction".
In this paper we want to advance on the basis of recent Direct Numerical
Simulation of model viscoelastic hydrodynamic equations \cite{97THKN,98DSB,00ACP,02ACP}.
Such DNS show unequivocally that drag reduction is reproduced by model
equations like the FENE-P model. From the
theoretical point of view this is significant, since it indicates that the
phenomenon is included in the solutions of the model equations. Understanding drag reduction
then becomes a usual challenge of theoretical physics.

Our thinking here is motivated in part by a recent analysis of the stability of
laminar channel flows subject to space dependent effective viscosity \cite{01GLP,02GLP}. 
It turned out the even small viscosity gradients can lead to a giant stabilization
of the most unstable modes, both for primary and secondary instabilities. 
In these cases one can understand the phenomenon completely by examining the
energy budget of the putative unstable modes and their interaction with the mean flow;
the most important observation had been that it is the the existence of
viscosity {\em gradients} positioned at a strategic distance from the wall
which is crucial for the existence of a large effect. 
It seems desirable to do something similar for the viscoelastic turbulent flows as well
(in which the space dependent effective viscosity arises due to differential
stretching of the polymers). But alas, in distinction from primary and secondary
instabilities, where it is obvious which are the relevant modes, for the turbulent flow
these are not known apriori. We therefore decided to first initiate a systematic study
of the empirical modes that are sustained in the turbulent flow, and then
to discuss their interaction with the mean flow and with the polymeric additive, 
their stabilization or destabilization
when we compare the viscoelastic to the Newtonian flow, and their energy budget.
In this paper we present the first results of this study.

We will demonstrate that we can determine with reasonable accuracy at least the
first thirty most energetic modes that are sustained in the turbulent
flow, for both the FENE-P and the Navier-Stokes equations (run at the same
friction Reynolds number, and see below for details). These modes
can be arranged in descending order according to their relative energy.
Unexpectedly we find that the nature of the most relevant modes is unchanged
in the two cases. On the other hand the energies associated with the modes
and their relative ordering are changed;
some modes that are energetic in one flow are strongly suppressed (their
energy decreases by a factor of 4) in the other flow, and vice versa.
Most importantly, the few most energetic modes of the viscoelastic
flow contain a lot more energy that the same number of most energetic
Newtonian modes. We propose therefore that drag reduction should
be understood by examining the dynamics and relative stability of
the energy containing modes rather than focusing only on the dissipative
end of the spectrum, as proposed for example in \cite{90Ge}.  

In Sects. \ref{Eqmot} A and B we summarize the FENE-P equations and the numerical approach.
In Sect. \ref{drag} we present the essential results regarding the observation of 
drag reduction. In Sect. \ref{empirical} we review the Karhunen-Lo\'eve method for 
determining the best empirical modes, and apply it to the problem at hand.
In Sect. \ref{results} we discuss the results, demonstrate the invariance of the modes,
and present the relative ordering. Sect. \ref{summary} is devoted to a discussion
of the findings and of the road ahead.
\section{Equations of Motion and Direct Numerical Simulations}
\label{Eqmot}
\subsection{The FENE-P model for dilute polymers}
\label{Fenep}
The addition of a dilute polymer to a Newtonian fluid gives rise
to an extra stress tensor $\B {\C T}(\B r,t)$ which affects the Navier-Stokes equations:
\begin{eqnarray}
\frac{\partial \B u}{\partial t}+(\B u\cdot \B \nabla) \B u&=&-\B \nabla p 
+\nu_{\rm s} \nabla^2 \B u +\B \nabla \cdot \B {\C T}\ , \nonumber\\
\B \nabla \cdot \B u &=& 0 \ . \label{Equ}
\end{eqnarray}
Here $\B u(\B r,t)$ is the solenoidal velocity field,  $p(\B r,t)$ is the pressure
and $\nu_s$ is the viscosity of the 
neat fluid. The additional stress tensor $\B {\C T}$ is not
known exactly, and needs to be modeled. There are a number of
competing derivations, see \cite{87BCAH,94BE}. In our work we adopt the FENE-P model that
is derived on the basis of a dumbbell model for the polymers  \cite{87BCAH,94BE} and is known to 
reproduce the phenomenon of drag reduction. In this model
$\B {\C T}$ is determined by the ``polymer conformation tensor" $\B R$ according to
\begin{equation}
\B {\C T}(\B r,t) = \frac{\nu_{\rm p}}{\tau_{\rm p}}\left[\frac{f(\B r,t)}{\rho_0^2} \B R(\B r,t) -\B 1 \right] \ .
\end{equation}
Here $\B 1$ is the unit tensor, $\nu_{\rm p}$ is a viscosity parameter, $\tau_{\rm p}$ is a relaxation time for the
polymer conformation tensor and $\rho_0$ is a parameter which in the 
derivation of the model stands for the rms extension of the
polymers in equilibrium. The function $f(\B r,t)$ limits the growth of the trace of
$\B R$ to a maximum value $\rho_{\rm m}$:
\begin{equation}
f(\B r,t) \equiv \frac{\rho_{\rm m}^2-\rho_0^2}{\rho_{\rm m}^2 -R_{\gamma\gamma}(\B r ,t)} \ .
\end{equation}

The model is closed by the
equation of motion for the conformation tensor which reads
\begin{eqnarray}
\frac{\partial  R_{\alpha\beta}}{\partial t}&+&(\B u\cdot \B \nabla) R_{\alpha\beta}
=\frac{\partial u_\alpha}{\partial r_\gamma}R_{\gamma\beta}
+R_{\alpha\gamma}\frac{\partial u_\gamma}{\partial r_\beta}\nonumber\\
&-&\frac{1}{\tau_{\rm p}}\left[ f(\B r,t)  R_{\alpha\beta} -\rho_0^2 \delta_{\alpha\beta} \right] \ .
\label{EqR}
\end{eqnarray}

The model is derived by assuming that the polymer can be characterized completely by
an end-to-end vector distance. Nevertheless the resulting equations could be written on the
basis of plausible arguments including up to quadratic terms in gradients and
available tensors. For our purposes the accuracy
of the model in reproducing quantitatively all the phenomena  of 
turbulence in viscoelastic fluids is not an issue of
concern. We are mainly interested in the fact that
these equations were simulated on the computer in a channel geometry and exhibited
the phenomenon of drag reduction as discussed below. Our aim is to understand drag reduction
within the FENE-P model.
\subsection{Direct Numerical Simulations}
\label{ss:dimensionless}
\begin{figure} 
\includegraphics[width=7.5cm]{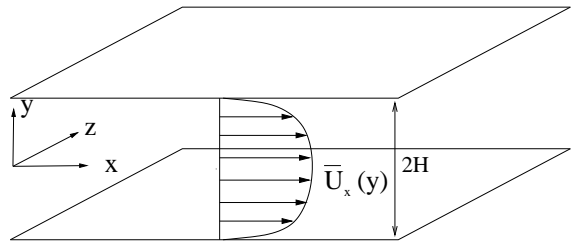}
\caption{ Geometry of the channel flow between two parallel planes
separated by $2H$ in the $y$ direction. The mean velocity
$\overline{\B U}(y)$ is oriented along the $x$ axis (streamwise
direction). The three-dimensional velocity fluctuations are
space-homogeneous in the $x-z$ plane, where $z$ is usually called
the span-wise direction. It is customary to use the Fourier
decomposition in this plane; $\B q =  ( q_x,  q_z )$ is the
corresponding two-dimensional wave vector. } \label{f:geometry}
\end{figure}

A simple flow geometry which exhibits the phenomenon of drag reduction is 
channel flow between two parallel
planes, separated by the distance $2 H$ in the $y$-direction, see
Fig.~\ref{f:geometry}. The computational domain is 
periodic in the two directions parallel to the wall (streamwise
$x$, spanwise $z$).
The Navier-Stokes equation (\ref{Equ}) is written in terms of
wall-normal component of velocity $u_y(\B r,t)$ and the vorticity $\omega_y(\B r,t)$, 
where $\B \omega = \B \nabla \times \B u$,
\begin{eqnarray}
  \label{NS2}
\frac{\partial u_y}{\partial t}  +  (\B u\cdot \B \nabla) u_y 
   &=& - \frac{\partial p}{\partial y}
+ \nu_s  \nabla^2 u_y + \nabla_{\beta}\C T_{\beta y}
  \nonumber \\
   \frac{\partial \omega_y}{\partial t}  + 
  \left( \B u  \cdot  \B \nabla\right)  \omega_y 
   &  =  & \left(\B \omega  \cdot  \B \nabla  \right)  u_y
  +  \nu_{\rm s}  \nabla^2  \omega_y   \nonumber  \\
 &  & +  \epsilon_{y \beta \gamma } \nabla_\beta  \nabla_{\delta} 
  \C T_{\delta \gamma} 
\end{eqnarray}
with $\epsilon_{\alpha  \beta  \gamma}$ the fully antisymmetric
Levi-Civita tensor. Given $u_y$ and $\omega_y$, the components of
the velocity field in the two directions parallel to the $x-z$
plane follow from the continuity equation and the definition of
$\omega_y$,
\begin{eqnarray}
 \label{kinematics}
 \B \nabla_\parallel  \cdot  \B u_\parallel   = 
 - \frac{\partial u_y}{\partial y} &, \qquad &
 \B \nabla_\parallel  \times  \B u_\parallel  =   \omega_y  ,
\end{eqnarray}
where the subscript $\parallel$ denotes projection of vectors on
the $x-z$ plane. The system of
Eqs.~(\ref{NS2},\ref{kinematics}) is one of the standard
formulations used for the direct numerical simulation of turbulent
channel flows, since the procedure yields a solenoidal velocity field
to machine accuracy.

In light of the periodicity in the $x-z$ plane, it is natural
to expand the planar components of the velocity field in Fourier modes.
For the wall-normal direction one uses Chebyshev polynomials.
The time stepping is carried out by a mixed
Crank-Nicolson/Runge-Kutta scheme for the viscous and the
nonlinear terms, respectively. The integration in the normal
direction is done by Chebyshev tau-method and a standard
de-aliasing technique is adopted for the nonlinear terms. 
The typical simulations have been performed on a computational
grid of $96 \times 129 \times 96$ nodes in a domain of dimensions
$2 \pi H \times 2 H \times 1.2 \pi H$. Turbulence is maintained by
enforcing the same constant pressure gradient for the corresponding
Newtonian and viscoelastic simulations.

In discussing the simulations, one notes that the channel half-height $H$
is only one of the parameters which sets up the external lengthscale. 
An additional important control parameter is the enforced pressure
gradient at the wall, which determines the friction. Denoting by pointed
brackets an average over time the friction parameter is defined as
\begin{equation}
 \label{eq:pressure}
\overline{\tau}_{\rm w} = H  \left\langle \frac{\partial
p}{\partial x} \right\rangle  \ .
\end{equation}
This is the basic control parameter of the flow. By considering
also the overall kinematic viscosity, $\nu_{\rm f} = \nu_{\rm s}
+ \nu_{\rm p}$, the traditional friction Reynolds number
is defined as
\begin{equation}
 \label{Reynolds}
{\rm Re}_\tau = \frac{u_\tau H }{\nu_{\rm f}} , \qquad 
u_\tau = \sqrt{\overline{\tau}_{\rm w}} \ ,
\end{equation}
where $u_\tau$ is the friction velocity. As customary in wall
turbulence, one may introduce the inner, or viscous, length scale
\begin{equation}
 \label{eq:viscous_scale}
\ell  = \frac{\nu_{\rm f}}{u_\tau} \ .
\end{equation}
The inner velocity scale $u_\tau$ has its counterpart bulk velocity
\begin{equation}
 U_0 = \frac{1}{2 H} \int dy  \langle u_x \rangle  \ .
\end{equation}
Correspondingly we have the outer and inner time scales $T_0=H/U_0$
and $\nu_{\rm f}/u_\tau^2$ respectively.

In the sequel all quantities are made
dimensionless unless stated otherwise. Those made dimensionless with respect to the
appropriate inner scale will be denoted by the superscript $+$;
no special symbol is used to denote normalization by an outer scale.

Finally there are parameters associated with the polymer. Foremost is the 
Deborah number which is defined as the
ratio of the relaxation time $\tau_{\rm p}$ and a typical time
scale of the fluid motion, i.e.
\begin{eqnarray}
   \label{eq:De}
    {\rm De} = \frac{\tau_{\rm p}}{T_0}  .
\end{eqnarray}
The parameter $r_{\rm p}=\nu_{\rm p}/\nu_{\rm s}$
measures the relative viscosity of the polymers with respect to
the Newtonian solvent, and the non-linear characteristics
of the spring is defined in terms of the ratio $\rho_{\rm
m}^2/\rho_0^2$.

In all our comparisons of Newtonian and viscoelastic simulations, the
friction Reynolds number is kept constant, at a typical value 
${\rm Re}_{\tau}=125$. The correspondence between the two flows is
obtained by fixing the computational domain, and choosing for
the Newtonian case a viscosity equal to the overall viscosity of
the solution. The parameters chosen for all the viscoelastic simulations are
${\rm De}=25$, $\eta_{\rm p}=0.1$ and $\rho_{\rm m}^2/\rho_0^2 =1000$. 
\subsection{Overview of drag reduction}
\label{drag}
In this subsection we review briefly the main results of the
present simulations which demonstrate the phenomenon of drag
reduction. For more detailed description, see \cite{02ACP}.
The analysis presented in the following sections employs the
very same DNS.

In comparing the viscoelastic and the Newtonian flows we maintain
the friction Reynolds number (\ref{Reynolds}) fixed. We reiterate that to achieve
this we need to choose the viscosity of the Newtonian flow properly, since
the viscoelastic wall stress contains a small viscoelastic
contribution,
\begin{equation}
\overline{\tau}_{\rm w}= \nu_{\rm s} \frac{d \overline{U}}{dy}+
\overline{\C T}_{y x}  \ .
\end{equation}
The component of the extra-stress $\C T_{y x}$
does not vanish on the average, contributing in our  simulation
about $10\%$ of the total drag.

\begin{figure} 
\includegraphics[width=9.5cm]{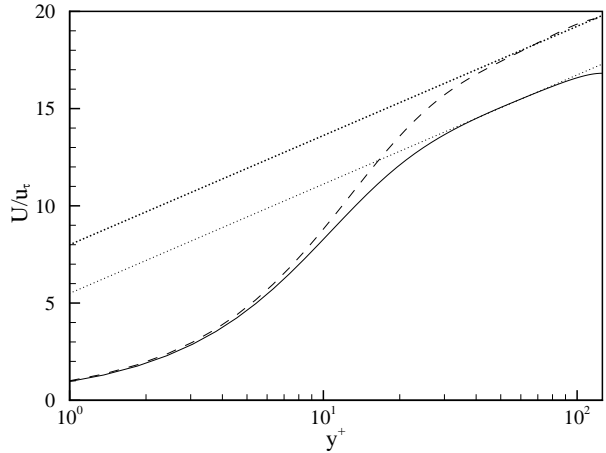}
\caption{ Mean velocity profiles for the Newtonian and for the
viscoelastic simulations with ${\rm Re}_\tau=125$.  Solid line:
Newtonian. Dashed line: Viscoelastic. The straight lines represent
the classical log-law, Eq.~(\ref{eq:log_law}).} \label{profile}
\end{figure}

The main observations regarding drag reductions are as follows:

(i) For a fixed mean pressure gradient at the wall, the viscoelastic flow
exhibits an increased flow rate through the channel, see Fig. \ref{profile}.
Shown there are the mean profiles in the streamwise direction, the other
means vanish by symmetry:
\begin{eqnarray}
  \label{eq:mean_profile}
  \overline{U}_x(y) & = & \overline{U}(y) \equiv  \langle u_x(\B r,t) \rangle  , \\
  \overline{U}_y(y) & = &
  \overline{U}_z(y)   =  0 \ . \nonumber
\end{eqnarray}
The increase in the throughput
entails an increased bulk Reynolds number,
\begin{equation}
 \label{eq:bulk_Re}
  {\rm Re}_{\rm b}  =  \frac{U_0 H}{\nu_{\rm f}} \ .
\end{equation}

\begin{figure} 
\includegraphics[width=9.8cm]{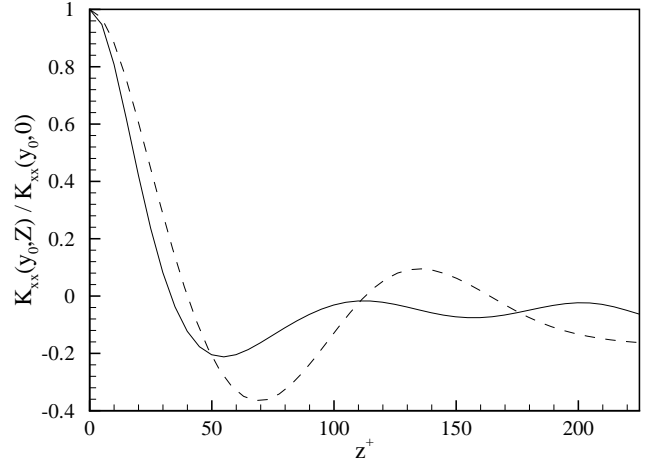}
  \caption{Two point span-wise correlation of the streamwise velocity component,
           $K_{x x}(y_0,Z)/K_{x x}(y_0,0)$: viscoelastic flow (dashed line),
            Newtonian flow
           (solid line). ${\rm Re}_\tau = 125$, $y_0^+ = 7$. }
\label{f:correlation}
\end{figure}

(ii)  Some typical length scales increase in the viscoelastic flow. The
data in  Fig.~\ref{profile} can be recast in terms of the log-law of the wall,
\begin{equation}
 \label{eq:log_law}
 \overline{U}^+  =  \frac{1}{k}  \log{y^+}  +  A  ,
\end{equation}
where $k \simeq 0.4$ is the Von Karman constant. The constant $A$
depends on the thickness of the buffer plus viscous sub-layers,
defined as the distance between the wall and the beginning of the
log-region. The log-law exists equally well for the viscoelastic
as for the Newtonian flow with the same $k$, but the numerical value
of the constant $A$ is substantially increased in the former. 
This thickening of the buffer
layer is directly related to the increased flow rate. The effect
of thickening buffer layer can be also measured by the increase of the
span-wise scale of the streamwise velocity fluctuations. Decomposing the
velocity field into a mean and a fluctuating part
\begin{equation}
  \label{eq:R_dec}
   u_\alpha (\B r,t)=\overline{U}_\alpha(y) +\tilde u_{\alpha}(\B r,t)  ,
\end{equation}
one introduces the correlation tensor
\begin{equation}
   \label{eq:corr_func}
   K_{\alpha \beta}(\B r,\B r' ) \equiv 
\langle \tilde u_\alpha(\B r,t)  \tilde u_\beta(\B r',t) \rangle \ . 
\end{equation}
In Fig.~\ref{f:correlation} we show $K_{x x}(\B r, \B r')$ for $\B r =
(x,  y, z)$, $\B r' = (x,  y, z + Z)$. [Note that by
homogeneity  $K_{x x } = K_{x x }(y, Z)$]. 
Fig.~\ref{f:correlation} demonstrated the increase in the  span-wise correlation length of the
streamwise velocity fluctuations which can be defined as
\begin{equation}
 \label{eq:corr_length}
 L_{x z}(y) \equiv \frac{1}{ K_{x x }(y,0)} 
 \int dZ  K_{x x }(y, Z) \ .
\end{equation}

(iii) Comparison of the time signals between the viscoelastic
and Newtonian cases shows an alteration of the characteristic
frequencies for the viscoelastic model. This corresponds quite
well to the experimental data, for example of Luchik et al.
\cite{Tiederman}, in which a decrease of the bursting frequency is
observed in drag reducing solutions. We do not reproduce these results
here.
\begin{figure}
\includegraphics[width=9.8cm]{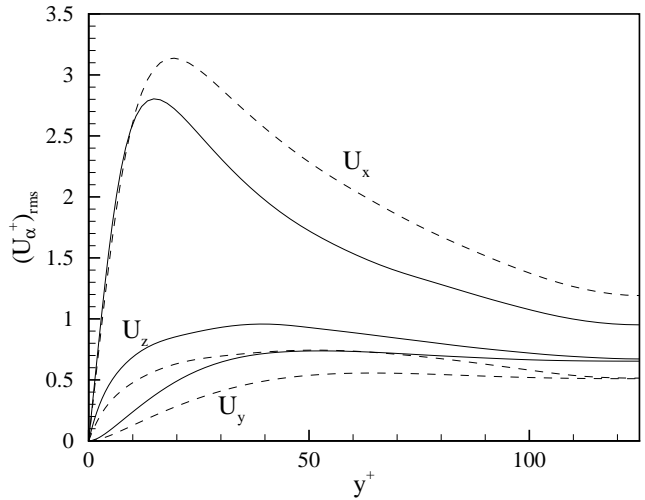}
\caption{Velocity fluctuations, normalized by friction velocity $u_\tau$,
         for the viscoelastic (dashed line) and the
         Newtonian flow (solid line).
         In both cases, $U_x$, $U_y$ and $U_z$ are given by
         solid dashed and dotted lines, respectively.}
\label{f:rms}
\end{figure}

(iv) Finally, the root mean square fluctuations change significantly 
as a function of $y$. Denoting
\begin{equation}
U_\alpha(y) \equiv \sqrt{\langle |\tilde u_\alpha(\B r,t)|^2 \rangle} \ ,
\end{equation}
we display in Fig. \ref{f:rms} the $y$ dependence of the three components $\alpha=x,y,z$.
The streamwise fluctuations are shown to increase with respect to the
corresponding Newtonian flow, while both the span-wise and the
wall-normal fluctuations decrease, in qualitative agreement with
available experimental results \cite{Tiederman}.

In conclusion, the FENE-P model is shown to exhibit the phenomenon
of drag reduction in close similarity to experimental observations.
In addition, DNS of this model provides complete information about
the velocity field and the covariance tensor field $\B R(\B r,t)$ as a function
of space and time. We thus feel confident that a sufficiently savvy
analysis of this model and its turbulent flow pattern should provide
insight into the mechanism of drag reduction.
\section{Analysis Using Empirical Modes}
\label{empirical}
In this section we present the analysis of the difference between
viscoelastic and Newtonian flows in term of the empirical modes
that are sustained in the turbulent flow. In choosing a method to extract
the modes we are led by the desire to find the ``best" modes,
and ``best" in our case will mean those that are most energetic. In fluid
mechanics the energy is quadratic in the field, so ``best" is related to 
closeness in the $L^2$ norm. As is well known, the standard method
to find best representation in $L^2$ norms is the Karhunen-Lo\'eve method.
The aim of the Karhunen Lo\'eve method is to provide a set of modes
that optimally decompose the field of interest, in our case the
velocity field averaged over time. The approach is guaranteed
to yield the best set of truncated modes, meaning that
the field cannot be approximated better (in $L^2$ norm) by any other set
of the same number of modes. The method had been applied to fluid mechanics in a 
number of contexts, see for example \cite{Siro1,Siro2,Holmes} for details and relevant
references. We first adapt the  Karhunen-Lo\'eve method to the present context, and then discuss the
results of the analysis.

The core object for the analysis is the simultaneous correlation 
function of the velocity fluctuations which was introduced already
in Eq.(\ref{eq:corr_func}). Due to stationarity this object
is time independent, and due
to homogeneity in the $x$ and $z$ direction, we can write
\begin{equation}
   \label{eq:corr_func2}
   K_{\alpha \beta} =
   K_{\alpha \beta} (x'-x,  z'-z;  y, y') \ .
\end{equation}
In the translationally invariant directions $x-z$ one cannot do better than
Fourier decomposition. Accordingly we consider the partly decomposed object
$\hat Q (\B q|y,y')$ defined as
\begin{eqnarray}
\hat Q (\B q|y,y') &\equiv& \frac{1}{N_\parallel} 
\sum_{\B r_\parallel}  K_{\alpha \beta} (x'-x,  z'-z;  y, y')
\\ &&\times \exp\{ -\imath [q_x (x'-x) + q_z (z'-z )]\} \ . \nonumber
\end{eqnarray}
We denoted the discrete two
dimensional wave-vector in the $x-z$ plane as $\B q = (q_x, q_z) = (\tilde{q}_x  2 \pi/
\Lambda_x, \tilde{q}_z  2 \pi/ \Lambda_z)$ where $\tilde{q}_x, 
\tilde{q}_z$ are integers. $N_\parallel = N_x  N_z$ and the sum
is taken over the discrete set of $x-z$ points with
$\B r_\parallel = (j_x \Lambda_x/N_x,  j_z \Lambda_z/N_z) $
the $x-z$ projection of $\B r$. 

The non-trivial empirical modes are obtained from the remaining dependence on
$y,y'$, for each given planar $\B q$. The Karhunen-Lo\'eve method consists of finding the 
eigenfunctions $\Psi_{\alpha\beta}(\B q,p|y)$ which solve the
eigenvalue equation 
\begin{equation}
   \label{eq:eigen}
     \int_{-1}^{1} \hat Q_{\alpha \beta}(\B q|y, y') \Psi_{\beta}(\B q, {\rm p}| y')  dy'  = 
     E(\B q, {\rm p})
   \Psi_{\alpha}(\B q,  {\rm p}|  y) \ .
\end{equation}
Here $E(\B q,{\rm p})$ is the energy associated with the mode $(\B q,
{\rm p})$, ordered in decreasing energy order,  with a given $\B q$ wave-vector 
in the wall parallel
directions, and p mode number in the $y$-direction, orthogonal to
the plane walls. Denoting by $\C E$ the total energy in the system,
we have
\begin{equation}
  \label{eq:tot_energy}
  \C E \equiv \int  d^3  \B r \frac{\tilde u^2}{2}  = \sum_{\B q, {\rm p}}  E(\B q, {\rm p})  .
\end{equation}
The modes labeled by $(\B q,  {\rm p})$ can be relabeled in
order of decreasing energy by introducing an index $n$. In this
notation the mode whose associated energy is $E_n = E(\B q,  {\rm
p})$ are denoted by
\begin{equation}
   \label{eq:global_modes}
   \Psi_\alpha(n|y)  \equiv  \Psi_\alpha(\B q_n,  {\rm p_n}| y)   ,
\end{equation}
where $(\B q_n,  {\rm p}_n)$ is the label of the $n^{\rm th}$ mode. It
is useful to introduce also a set of fields indexed by $n$,  
\begin{equation}
 \label{eq:physical_modes}
 \Phi_\alpha(n| \B r)  = 
 \Psi_\alpha(n|y)   \exp\left[\imath  (\B q_n  \cdot  \B r_\parallel) 
\right]  \ .
\end{equation}
These function are orthogonal in the sense
\begin{equation}
 \label{eq:norm}
\frac{1}{2 N_\parallel} \sum_{\B r_\parallel }\int_{-1}^1 dy
   \Phi_\alpha(n| y, \B r_\parallel ) 
  \Phi^*_\alpha(n'| y, \B r_\parallel )  = 
 \delta_{n n'}  \ .
\end{equation}
The fluctuation velocity field can be now expanded in terms of these fields,
\begin{equation}
\label{eq:expansion} \tilde u_\alpha(t,\B r)  =  \sum_n 
a_n(t) \Phi_\alpha (n| \B r) . 
\end{equation}
The main advantage of the Karhunen-Lo\'eve method is that any finite
truncation of this expansion can be shown to yield a best approximation
for the velocity field (in the $L^2$ norm). Moreover, due to the symmetry of
the correlation matrix the modes are orthonormal (after suitable normalization),
and correlation functions in the basis of these modes are diagonal:
\begin{equation}
 \label{eq:corr_diag} 
\langle a_n a^*_{n'} \rangle =  E_n \delta_{n n'} \ .
\end{equation}
Similarly one can have an optimal representation of the correlation matrix itself as 
\begin{equation}
   \label{eq:corr_func3}
   K_{\alpha \beta} = \sum_n  E_n  \Psi_\alpha(n|y)\Psi_\beta(n|y')
\exp\left[ \imath \B q_n \cdot (\B r_\parallel -
\B r'_\parallel)\right]  \ . 
\end{equation}

In analyzing the DNS the modes are determined by using an expansion in 
terms of Chebyshev polynomials $T_k(y)$ in the $y$
direction. Denote $y_j$ the j$^{th}$ node in the list of $N_y$
Chebyshev nodes,
\begin{equation}
  \label{eq:CH_nodes}
  y_j= - \cos{\left(j\frac{\pi}{N_y} \right)}
  \qquad j=1,  \ldots,  N_y ,
\end{equation}
and express
\begin{equation}
   \label{eq:Cheb_exp}
   \Psi_\alpha(\B q,{\rm p}| y)  =
   \sum_{k=0}^{N_y-1}
   \hat{\Psi}_\alpha(\B q,{\rm p}| k) T_k(y)  \ .
\end{equation}
Then Eq.~(\ref{eq:eigen}) is discretized as
\begin{eqnarray}
   \label{eq:eigen_discr}
    \sum_{k=0}^{N_y-1}&& A_{ \alpha \beta}(\B q| j,k) 
     \hat{\Psi}_\beta(\B q,{\rm p}|k)     \nonumber \\
       &&= E(\B q,  {\rm p})
   \sum_{k'=0}^{N_y-1} 
     \hat{\Psi}_\alpha(\B q,  {\rm p}| k') T_{k'}(y_j) \ .
\end{eqnarray}
where
\begin{equation}
   \label{eq:eigen_mat}
   A_{ \alpha \beta}(\B q|j, k) \equiv 
   \sum_{j'=1}^{N_y} 
   \hat Q_{\alpha \beta}(\B q|y_j, y_{j'})  T_k(y_{j'}) w_{j'}
  \ .
\end{equation}
with $w_j$ the integration weights.

For a given $\B q$, Eq.~(\ref{eq:eigen_discr}) results in a linear 
algebraic eigenvalue problem of order  $(N_y 
\times  N_y)$, with the
p$^{th}$ eigenvector given by $ \hat{\Psi}_\alpha(\B q,  {\rm p}|
 k)$, $k=0,  \ldots,  N_y$ and the corresponding eigenvalue
$E(\B q,{\rm p})$. Since we need to solve for every discrete
wave-vector $\B q$, there exist in total $N_x  N_z$ eigenvalue problems
to be solved in order to determine the full set of eigenfunctions.
From the discrete eigenvectors, the corresponding discrete modes
are constructed according to the discrete analog of
Eq.~(\ref{eq:physical_modes}),
\begin{equation}
  \label{eq:physical_modes_discr}
 \Phi_\alpha(\B q,{\rm p}|  \B r_\parallel, y_j)  = 
 \Psi_\alpha(\B q, {\rm p}|  y_j) \exp\left[\imath (\B q_n  \cdot  \B r_\parallel) 
\right] \ , 
\end{equation}
where Eq.~(\ref{eq:Cheb_exp}) is used to evaluate $\Psi_\alpha$ at
the Chebyshev nodes. The discrete modes are normalized according
to the discrete version of Eq.~(\ref{eq:norm}),
\begin{equation}
 \label{eq:norm_discr}
 \frac{1}{2 N_\parallel} \sum_{j=1}^{N_y}  \sum_{\B r_\parallel} 
  \Phi_\alpha(\B q,{\rm p}| \B r_\parallel,  y_j) 
  \Phi^*_\alpha(\B q,  {\rm p}| \B r_\parallel,  y_j) 
   w_j  =  1 \ .
\end{equation}

In practice we ran DNS with and without polymers for $5000$ large-eddy
turnover times $T_0$, see Sec.~\ref{ss:dimensionless}, in
statistically stationary conditions. The time step used for
time-advancement, in terms of viscous units $\nu_{\rm
f}/u^2_\tau$, Sec.~\ref{ss:dimensionless}, is $Dt^+=0.05$. Since
the same pressure gradient is enforced in the two cases, the
friction velocity is identical, while the bulk velocity increases
by  $24 \%$ in the viscoelastic simulation. This corresponds to
a reduction of the large-eddy turnover time, $T_0$.

We have collected $N_{\rm fields}$ fields, $ N_{\rm fields} \simeq 100$, 
displaced in time by $50 ~T_0$. These fields were used to construct the discrete
correlation function, Eq.~(\ref{eq:corr_func}), at the nodes of
the computational grid. Eigenvalues and eigenvectors of its
discrete Fourier transform in the wall parallel directions has then been evaluated, according to
Eq.~(\ref{eq:eigen_discr}). The modes corresponding
to the computed eigenvectors have been arranged in order of
decreasing eigenvalue. From Eq.~(\ref{eq:norm_discr}), the
eigenvalues correspond to the average energies in the considered
modes, so that the chosen arrangement is in fact an ordering in
terms of decreasing energy content. When needed, the index of the
energetic ordering of the Newtonian and viscoelastic simulations will be denoted by
$n_{_{\rm N}}$ and $n_{_{\rm VE}}$ respectively, dropping the subscript 
when $n$ refers to both cases.

\section{Analysis and results}
\label{results}
\subsection{The dominant modes are approximately invariant}

The first discovery in the study of the empirical modes was admittedly 
surprising for the present authors,
and in hindsight very serendipitous for the discussion of drag reduction.
We found that {\em the dominant modes are approximately invariant}. In other
words, the same modes that carry a sizable fraction of the energy of the
viscoelastic flow appear essentially unchanged in the Newtonian flow,
having practically the same spatial $y$-dependence.
We will first demonstrate this surprising finding, and later focus
on the difference between the two flows.

To discuss meaningfully the correspondence between the empirical modes
in the  two cases we seek a criterion of 
matching the viscoelastic modes to the corresponding Newtonian modes. This
can be done easily, since both sets of modes form a complete
orthonormal basis for solenoidal fields in the same domain. Each
viscoelastic mode can then be expanded in terms of the Newtonian
set,
\begin{equation}
 \label{eq:exp_VE_N}
 \Psi_\alpha^{\rm VE}(\B q, \, {\rm p}_{_{\rm VE}}|y)
 =  \sum_{{\rm p}_{_{\rm N}}} \, A(\, \B q, {\rm p}_{_{\rm VE}}| \,
{\rm p}_{\rm N}) \,
 \Psi_\alpha^{\rm N}(\B q, \, {\rm p}_{_{\rm N}}| \, y)  \ .
\end{equation}
The complex amplitude $ A(\, \B q, {\rm p}_{_{\rm VE}}|{\rm
p}_{_{\rm N}})$ is given by the projection of the viscoelastic mode
${\rm p}_{_{\rm VE}}$ on Newtonian mode ${\rm p}_{_{\rm N}}$, with
identical wall-parallel wave-vector, $\B q$:
\begin{eqnarray}
 \label{eq:amp_VE_N}
&& A(\B q, {\rm p}_{_{\rm VE}}| {\rm p}_{_{\rm N}})=  \\
&&\frac{1}{2 N_y} \sum_{j=1}^{N_y} \,
 \Psi_\alpha^{\rm VE}(\B q, \, {\rm p}_{_{\rm VE}}| y_j)  \,
 \Psi_\alpha^{* \rm N}(\B q, \, {\rm p}_{_{\rm N}}|  y_j) w_j \ .
 \nonumber
\end{eqnarray}

\begin{figure} 
\includegraphics[width=8.5cm]{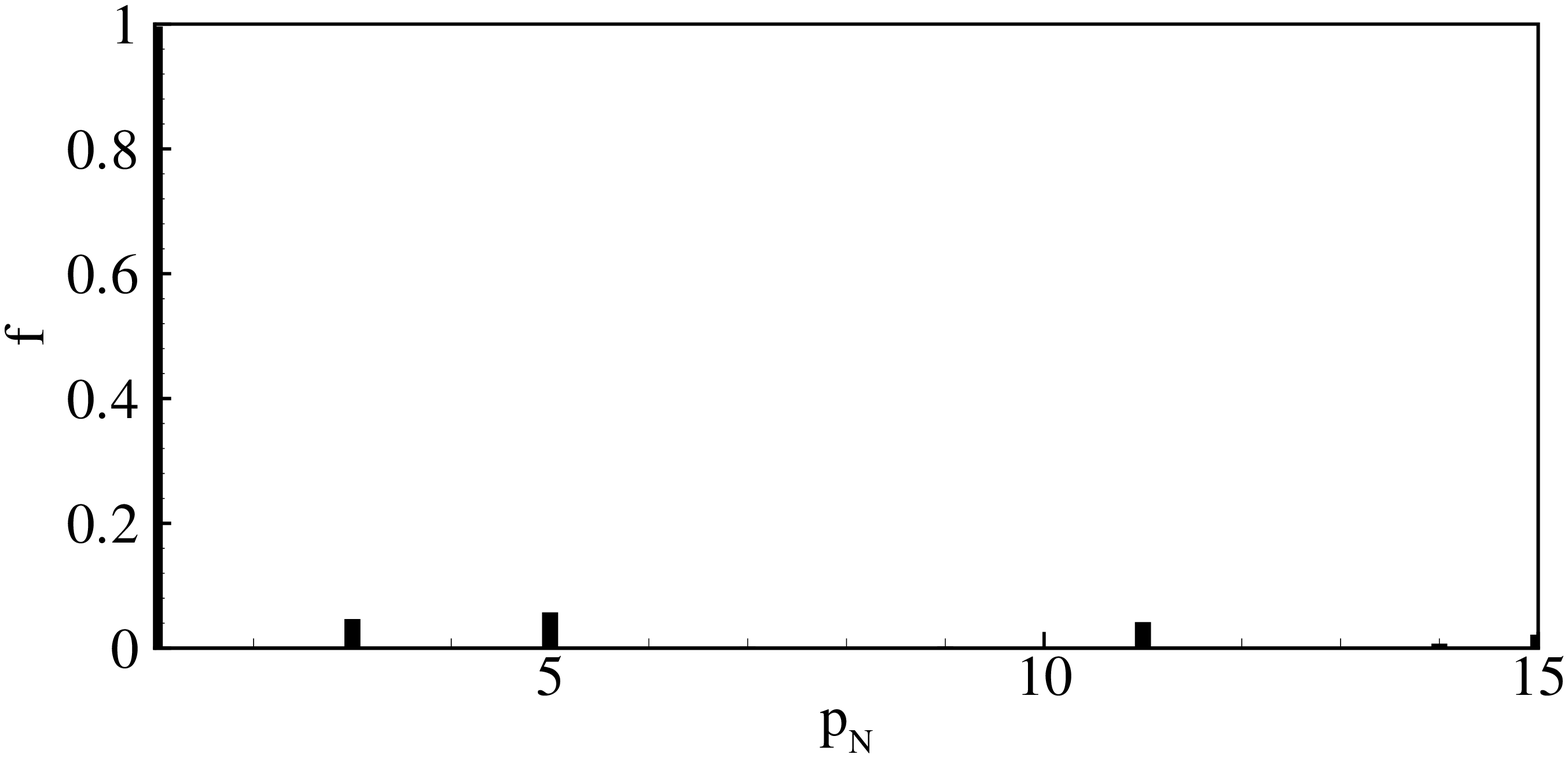}
\includegraphics[width=8.5cm]{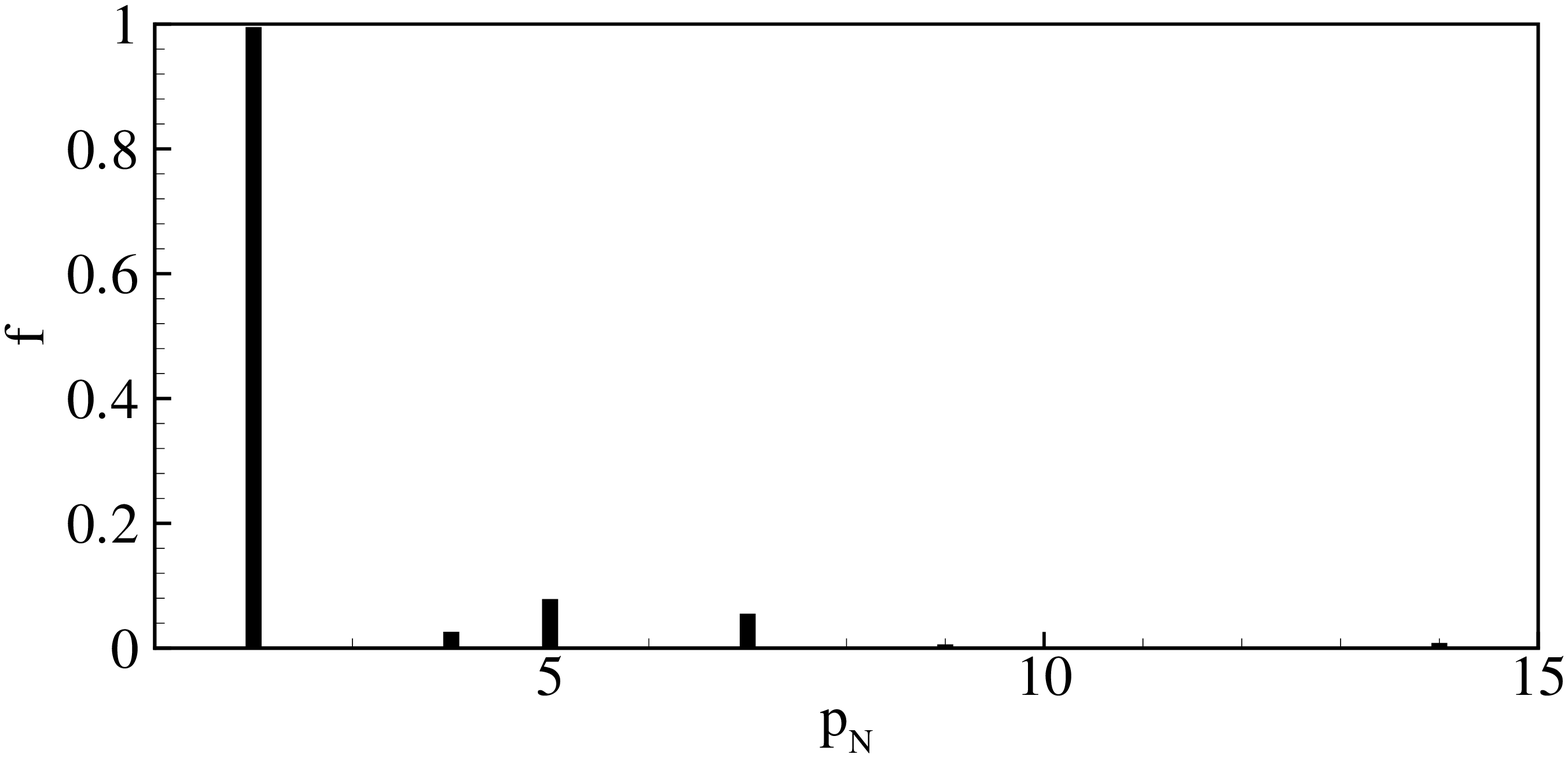}
\includegraphics[width=8.5cm]{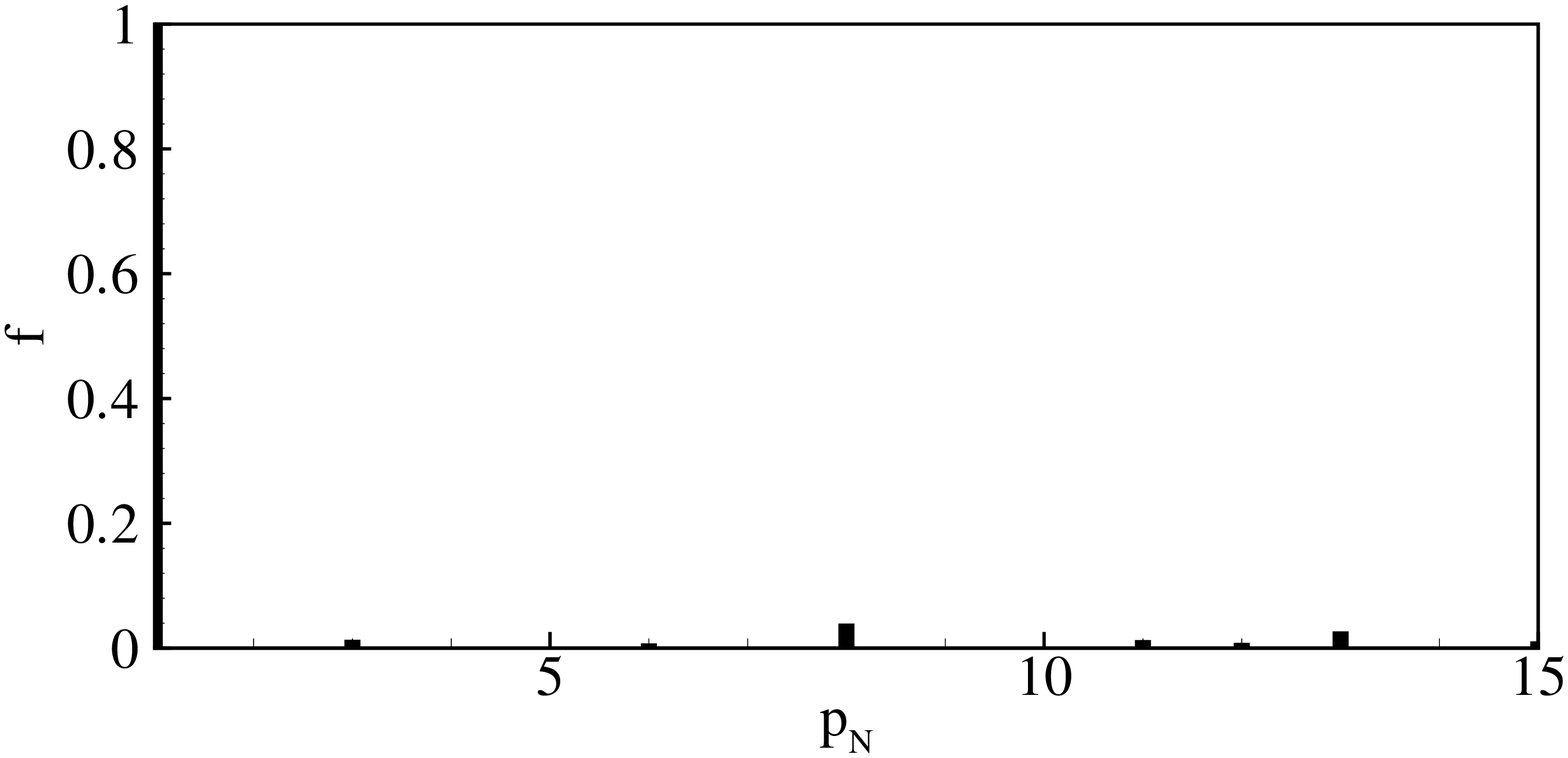}
\caption{Projections of the three most energetic viscoelastic modes on
the basis of the Newtonian best modes with the same $\tilde \B q$, ${\rm
Re}_\tau=125$. Panel a: the most energetic Viscoelastic mode,
$\tilde \B q = (0, 3)$, ${\rm p} = 1$. This modes fits very well the most 
energetic mode of the Newtonian flow which is the same, (0,3,1). Panel b: the second most
energetic Viscoelastic mode, $\tilde \B q = (0,2)$, ${\rm p} = 1$. This
modes fits very well the sixth most energetic mode of the Newtonian flow
(0,2,2).
Panel c: the third most energetic Viscoelastic mode, $\tilde \B q = 
(0,1)$, ${\rm p} = 1$. This mode fits well the fourth most energetic
Newtonian mode (0,1,1)} \label{f:Proj_VE_1-3}
\end{figure}

We find that all the most energetic modes of the viscoelastic
flow have one amplitude whose magnitude is close to unity. 
For example we plot
the absolute magnitude $|A(\B q, {\rm p}_{_{\rm VE}}|{\rm p}_{_{\rm N}})|$ 
as a function of ${\rm p}_{_{\rm N}}$ in Fig.~\ref{f:Proj_VE_1-3} for
the first three most energetic viscoelastic modes. Each of these modes displays an
amplitude maximum very close to unity, implying that it matches well a single Newtonian mode.
This procedure furnishes a 
correspondence between viscoelastic and Newtonian modes: a
given viscoelastic mode corresponds to the
Newtonian mode for which the amplitude $|A(\B q, {\rm p}_{_{\rm VE}}|{\rm p}_{_{\rm N}})|$
is maximal. This
unambiguously associates a single Newtonian mode to each
viscoelastic mode. The value of the maximal amplitude, hereafter
called {\sl matching parameter}, gives a quantitative estimate of
the difference in shape between two corresponding modes: A
matching parameter equal to one implies absolute identity
between the two modes in the energy norm. The correspondence in
the spatial  structure can then be verified by direct inspection.
For instance, matching modes for the cases discussed in
Fig.~\ref{f:Proj_VE_1-3}, where the matching parameter is well
above $0.9$, are almost indistinguishable. Physically, the
matching parameter represents the fraction of the energy in the
viscoelastic mode that is ascribed to the matching Newtonian mode.

\begin{figure} 
\includegraphics[width=8.5cm]{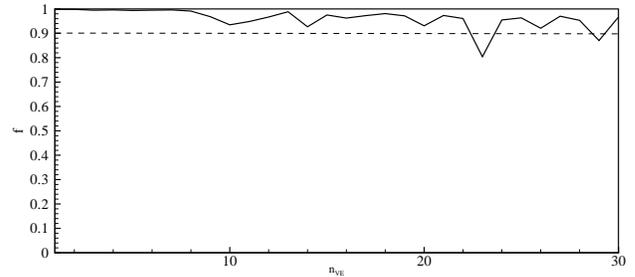}
\caption{Matching parameter of the
first $30$ most energetic viscoelastic modes with {\sl
corresponding} Newtonian modes, ${\rm Re}_\tau=125$. }
\label{f:Correlation_VE_N}
\end{figure}

Fig.~\ref{f:Correlation_VE_N} shows the matching parameter for the
first thirty most energetic viscoelastic modes ($n_{_{\rm VE}}\le 30$).
All the VE-modes except the 23rd and 29th have
a matching parameter above $0.9$. We conclude that at least
as far as the most energetic modes are concerned, the spatial
structure of the modes is almost not altered by the
polymer. Loosely speaking this can be expressed by saying  that
the modes are approximately invariant, i.e. fixed in shape.
{\em For practical purposes the difference between the Newtonian
and viscoelastic empirical modes can be safely disregarded}.
Next we discuss what changes from Newtonian to viscoelastic turbulence.
\subsection{The energy and the relative ordering change}

In light of the first discovery the second may be already anticipated:
although the dominant modes hardly change, their energies and relative
energy ordering change very significantly.
We propose that understanding the change of energies and relative ordering
of the modes will take us a long way in understanding drag reduction.
We begin by comparing the most dominant modes in the two respective flows.
Each empirical mode is identified by three numbers, $ (\tilde q_x, \tilde q_z)$, and p,
where as explained above, ${\rm p}$ corresponds to the
index  of the energetic ordering for fixed wave-vector $\B q$.
\subsubsection{The most dominant mode}
\begin{figure}
\includegraphics[width=7.5cm]{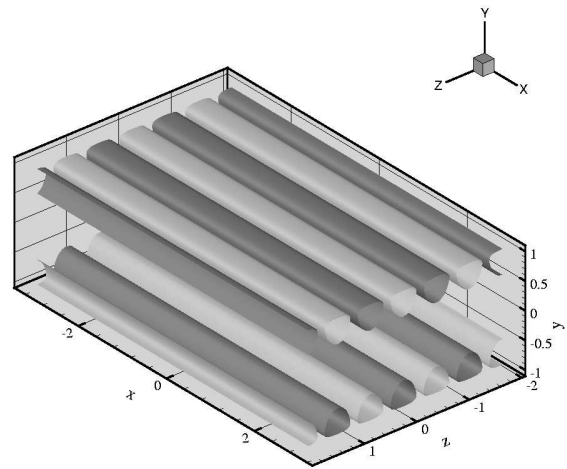}
\caption{Iso-surface of the streamwise component of the first most
energetic mode, $\Phi_1(n=1| \B r)$ which is the same for the Newtonian
and the  viscoelastic flows, i.e. the mode $\tilde \B q  = (0, 3)$, ${\rm p}= 1$.
In both cases, ${\rm Re}_\tau = 125$. Note that this mode is a {\rm non-propagating} roll mode.}
\label{f:non_propagating_mode1}
\end{figure}
In Fig. \ref{f:non_propagating_mode1} we present
pictorially the mode (0,3,1) which is the most energetic for
both the
Newtonian and viscoelastic simulation. Its amplitude $a_1(t)$,
\begin{equation}
   \label{eq:amplitude_m1}
   a_1(t) = \frac{1}{2 N_y} \sum_{j=1}^{N_y} \, \sum_{\B r_\parallel} \,
   \tilde \B u_\alpha(t, y_j, \, \B r_\parallel) \,
   \Phi^*_\alpha(n = 1|y_j, \, \B r_\parallel)w_j  \ ,
\end{equation}
is real in both flows, implying that the mode does not propagate, neither in the
$x$ nor in the $z$ direction. Moreover, for this particular mode
$q_x = 0$, i.e. the field $\Phi_\alpha(n=1|\B r)$ is constant
in $x$. It belongs to the class of non-propagating roll-modes
discussed in \cite{Siro2}. The figure presents an
iso-surface of the streamwise component of the velocity field
associated with the mode, $\Phi_x(n=1 |\B r) = {\rm const}$.
The span-wise wavelength of the mode is $\lambda_z = \Lambda_z
/\tilde{q}_z = 2 \pi H/3$, and it appears to be confined relatively
close to the channel walls. 

\begin{figure} 
\includegraphics[width=7.5cm]{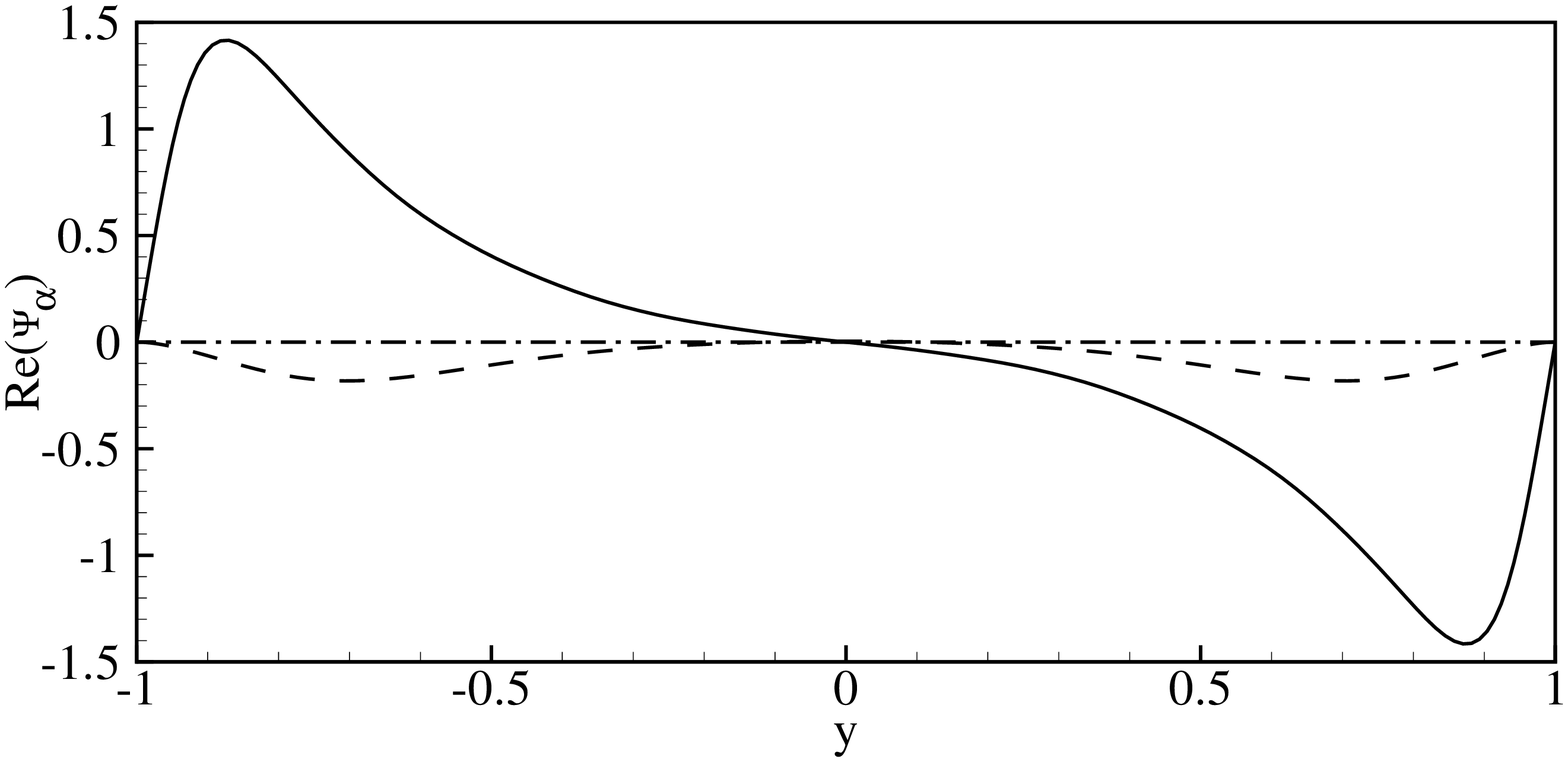}
\includegraphics[width=7.5cm]{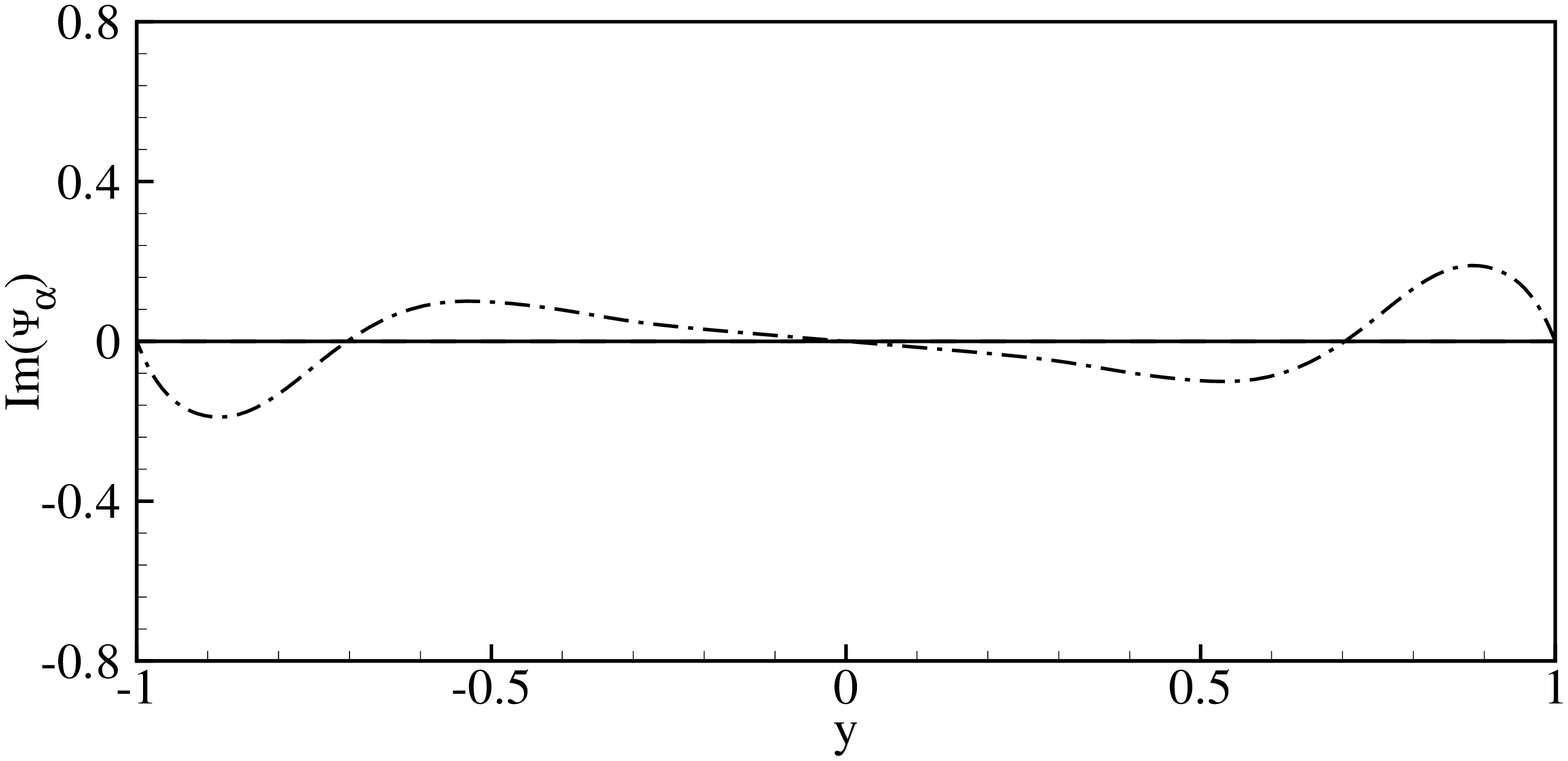}
\caption{Portrait of the most energetic Newtonian mode, $(0,3,
1)$: Upper panel - real part of the fluctuating velocity
profile  $\B \Psi(1| y)$, lower panel - imaginary part of
$\B \Psi(1| y)$. Solid lines - $\Psi_x(1| y)$, dashed lines -
$\Psi_y(1| y)$ and dash-dotted lines - $\Psi_z(1|y)$.} \label{f:Pass_N_1}
\end{figure}

A further pictorial presentation of the Newtonian mode (0,3,1) 
is shown in Fig.~\ref{f:Pass_N_1}, where the real part of the mode
(top panel) and imaginary part (bottom panel) are plotted 
separately. From the plots
it is apparent that the mode is more or less localized in the
vicinity of the walls, as already commented. The phase difference
between the various components is also worth mentioning. Comparing
the top and the middle panels, the stream-wise and span-wise
components, $\Psi_x$ and $\Psi_z$ are real for all $y$, while the
wall-normal component, $\Psi_y$, has a constant phase lag of
$\pi/2$ with respect to the other two, i.e. it is purely
imaginary. 
\begin{figure} 
\includegraphics[width=7.5cm]{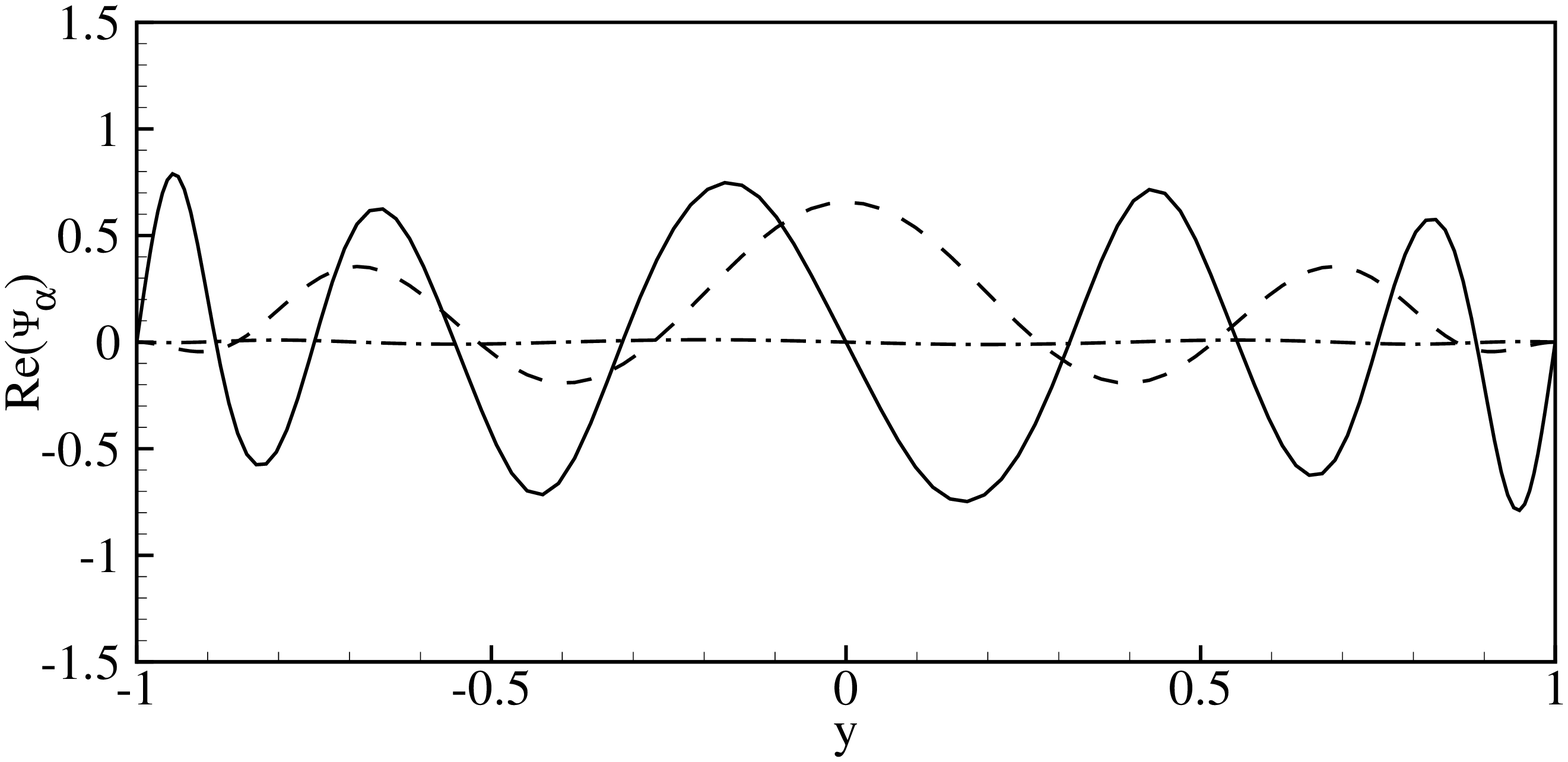}
\includegraphics[width=7.5cm]{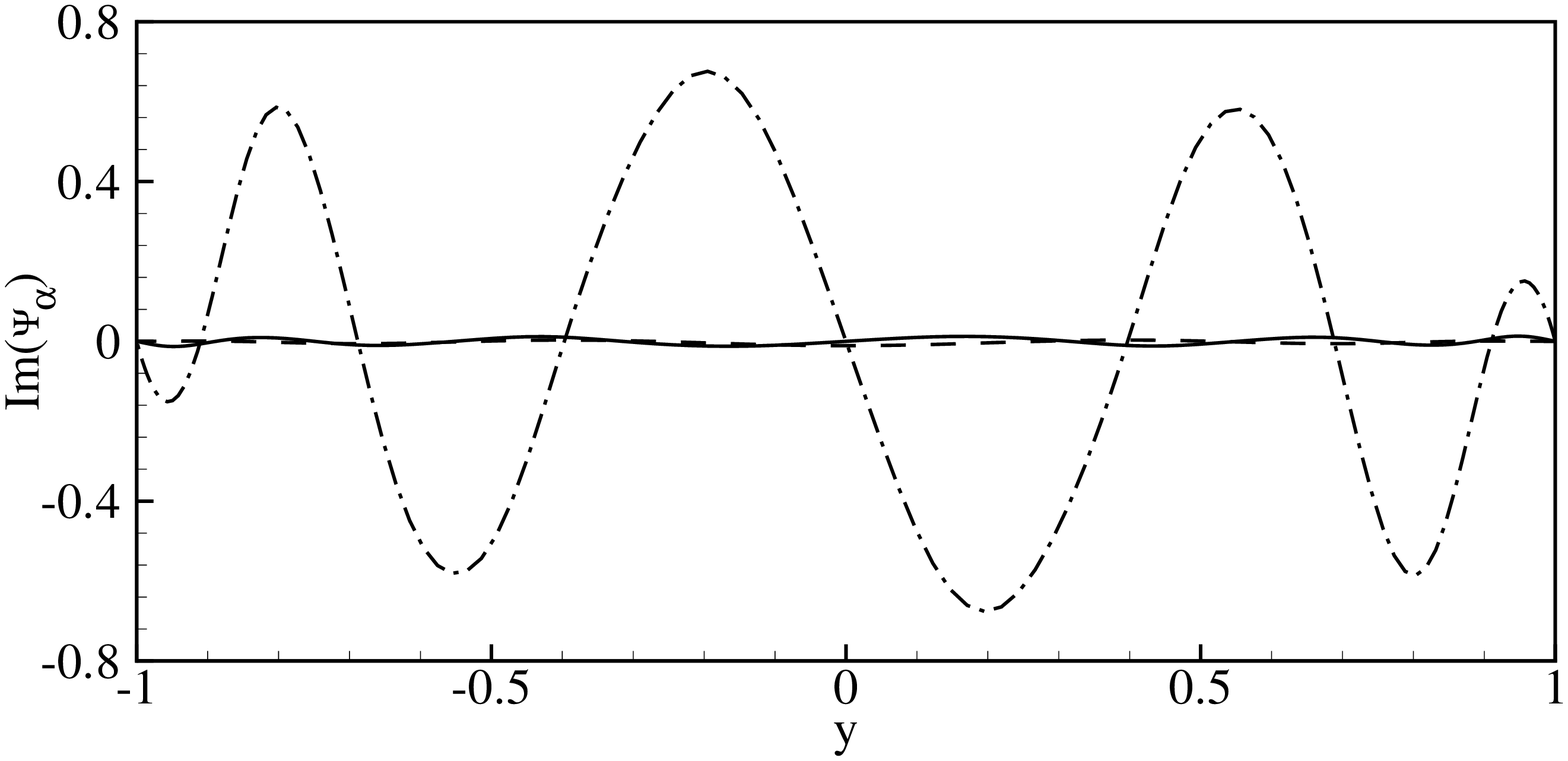}
\caption{Portrait of a typical viscoelastic high order mode, $(0,
3,15)$. Notations as in Fig.~\ref{f:Pass_N_1}. }
\label{f:Pass_VE_ho}
\end{figure}
We note that although the first, most dominant mode, is the same
for the two flows, the actual energy associated with this mode is
twice larger in the viscoelastic mode. This is typical for all
the leading modes in the viscoelastic flow as compared with the
Newtonian flow; this central point of distinction between the 
two flows will be addressed in the next sub-section. The agreement
between the leading modes ends with the first one; the majority  of higher
modes change in relative position in the energy descending ordering,
as explained below. Note that all the seven leading modes in both flows 
are associated with $q_x=0$. This is surely due to the relatively
short channel length in our computation. It is known that in longer
channels the ``roll" modes become oscillatory modes with a finite
value of $q_x$.

For the sake of completeness we present in
Fig.~\ref{f:Pass_VE_ho} the portrait of a higher order
viscoelastic mode, namely $\tilde \B q = (0, \, 3)$, ${\rm p} = 15$. 
We see that for this mode $\Psi_z$ is essentially purely imaginary, whereas
$\Psi_x$ and $\Psi_y$ are essentially purely real. In fact, this
particular mode is very low in energy, and we present it just to demonstrate
that the numerics is still not noisy even for rather low energy modes. 
\subsubsection{Sub-dominant modes and energy redistribution}
The full comparison between the first 30 modes of the Newtonian
and viscoelastic flows can be seen in Table 1, in which these
modes are listed together with their energy (in percent of the
total sum of energies). Also included in the table is the cumulative
sum of energies up to the mode listed. 
\begin{table}
\begin{center} 
\begin{tabular}{||r| ccc||c|ccc||}
\multicolumn{1}{||c}{} & 
\multicolumn{3}{c||}{Newtonian} &
\multicolumn{1}{c}{} & 
\multicolumn{3}{c||}{Polymers} \\
\multicolumn{1}{||c}{} & 
\multicolumn{3}{c||}{${\rm Re}_{\tau}=125$} &
\multicolumn{1}{c}{} & 
\multicolumn{3}{c||}{${\rm Re}_{\tau}=125$} 
 \\ \cline{1-4} \cline{5-8}
       & Mode &Energy & Sum & & Mode & Energy & Sum \\ 
$n_{_{\rm N}}$  & {\sl ($\tilde{q}_x$,$\tilde{q}_z$, ${\rm p}$)} & $(\%)$&$(\%)$& $n_{_{\rm N}}$ & {\sl ($\tilde{q}_x$,$\tilde{q}_z$, ${\rm p}$)} &$(\%)$ & $(\%)$\\ \hline
$ 1$ & $(0,3,1)$ & $   3.893 $ & $ 3.89 $& $  1$ & $(0,3,1)$ & $   8.625 $ & $    8.63 $ \\
$ 2$ & $(0,3,2)$ & $   3.882 $ & $ 7.77 $& $  6$ & $(0,2,1)$ & $   6.760 $ & $   15.39 $ \\
$ 3$ & $(0,2,1)$ & $   3.700 $ & $11.47 $& $  4$ & $(0,1,1)$ & $   6.580 $ & $   21.97 $ \\
$ 4$ & $(0,1,1)$ & $   3.218 $ & $14.69 $& $  3$ & $(0,2,2)$ & $   5.352 $ & $   27.32 $ \\
$ 5$ & $(0,1,2)$ & $   3.027 $ & $17.72 $& $  5$ & $(0,1,2)$ & $   4.780 $ & $   32.10 $ \\
$ 6$ & $(0,2,2)$ & $   2.828 $ & $20.55 $& $  2$ & $(0,3,2)$ & $   4.382 $ & $   36.48 $ \\
$ 7$ & $(0,4,1)$ & $   2.082 $ & $22.63 $& $  7$ & $(0,4,1)$ & $   2.477 $ & $   38.96 $ \\
$ 8$ & $(0,4,2)$ & $   1.624 $ & $24.25 $& $  9$ & $(1,2,1)$ & $   2.095 $ & $   41.05 $ \\
$ 9$ & $(1,2,1)$ & $   1.568 $ & $25.82 $& $  8$ & $(0,4,2)$ & $   1.934 $ & $   42.99 $ \\
$10$ & $(1,3,1)$ & $   1.364 $ & $27.19 $& $ 16$ & $(1,2,2)$ & $   1.610 $ & $   44.60 $ \\
$11$ & $(1,4,1)$ & $   1.276 $ & $28.46 $& $ 18$ & $(1,1,1)$ & $   1.584 $ & $   46.18 $ \\
$12$ & $(1,3,2)$ & $   1.265 $ & $29.73 $& $ 12$ & $(1,3,1)$ & $   1.527 $ & $   47.71 $ \\
$13$ & $(0,5,1)$ & $   1.226 $ & $30.95 $& $ 14$ & $(1,1,2)$ & $   1.431 $ & $   49.14 $ \\
$14$ & $(1,1,1)$ & $   1.179 $ & $32.13 $& $ 10$ & $(1,3,2)$ & $   1.336 $ & $   50.47 $ \\
$15$ & $(1,4,2)$ & $   1.112 $ & $33.24 $& $ 17$ & $(0,5,1)$ & $   1.282 $ & $   51.76 $ \\
$16$ & $(1,2,2)$ & $   1.081 $ & $34.32 $& $ 45$ & $(0,0,1)$ & $   1.275 $ & $   53.03 $ \\
$17$ & $(0,5,2)$ & $   1.014 $ & $35.34 $& $ 19$ & $(0,1,3)$ & $   1.272 $ & $   54.30 $ \\
$18$ & $(1,1,2)$ & $   1.010 $ & $36.35 $& $ 11$ & $(1,4,1)$ & $   1.036 $ & $   55.34 $ \\
$19$ & $(0,1,3)$ & $   1.000 $ & $37.35 $& $ 15$ & $(1,4,2)$ & $   1.035 $ & $   56.37 $ \\
$20$ & $(1,5,1)$ & $   0.817 $ & $38.17 $& $ 21$ & $(0,1,4)$ & $   1.015 $ & $   57.39 $ \\
$21$ & $(0,1,4)$ & $   0.804 $ & $38.97 $& $ 13$ & $(0,5,2)$ & $   0.967 $ & $   58.36 $ \\
$22$ & $(1,5,2)$ & $   0.792 $ & $39.76 $& $ 43$ & $(0,0,2)$ & $   0.801 $ & $   59.16 $ \\
$23$ & $(2,1,1)$ & $   0.620 $ & $40.38 $& $ 20$ & $(1,5,1)$ & $   0.636 $ & $   59.79 $ \\
$24$ & $(2,2,1)$ & $   0.586 $ & $40.97 $& $ 22$ & $(1,5,2)$ & $   0.620 $ & $   60.41 $ \\
$25$ & $(2,3,1)$ & $   0.562 $ & $41.53 $& $ 37$ & $(1,1,3)$ & $   0.614 $ & $   61.03 $ \\
$26$ & $(1,6,1)$ & $   0.561 $ & $42.09 $& $131$ & $(0,0,3)$ & $   0.571 $ & $   61.60 $ \\
$27$ & $(0,6,1)$ & $   0.554 $ & $42.64 $& $ 31$ & $(1,2,3)$ & $   0.536 $ & $   62.13 $ \\
$28$ & $(2,2,2)$ & $   0.519 $ & $43.16 $& $ 68$ & $(0,1,5)$ & $   0.495 $ & $   62.63 $ \\
$29$ & $(1,6,2)$ & $   0.512 $ & $43.68 $& $ 47$ & $(1,1,4)$ & $   0.493 $ & $   63.12 $ \\
$30$ & $(2,3,2)$ & $   0.511 $ & $44.19 $& $ 41$ & $(0,2,3)$ & $   0.489 $ & $   63.61 $ \\
\end{tabular}
\end{center}
\caption{Energy content of the first 30 eigenfunctions of the channel flow with and without
polymers. Re$_\tau = 125.$ 
}
\label{t:Table1}
\end{table}

To understand the Table we recall that for each value of $\B q$ the modes are
ordered by their energy and labeled by p. The over-all energy label is $n_{_{\rm N}}$ 
or $n_{_{\rm VE}}$. In the columns of the viscoelastic modes, instead of
writing $n_{_{\rm VE}}$ in obvious increasing order, we provided the value
of the Newtonian index $n_{_{\rm N}}$ for which the viscoelastic mode has
the highest matching parameter. Thus for example the second leading viscoelastic
mode with $n_{_{\rm VE}}=2$ matches almost perfectly the sixth Newtonian
mode $n_{_{\rm N}}=6$, etc.
\begin{figure} 
\includegraphics[width=8.5 cm]{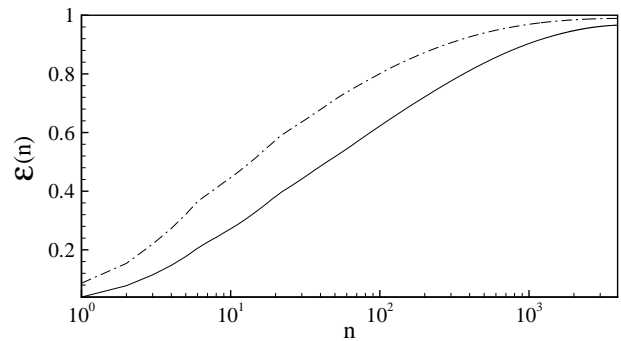}
\caption{ Energy sum $\C E(n) = \sum_{j=1}^n E(j)$ 
for the Newtonian (solid lines) and for the visco-elastic (dashed lines)
flow. The sums are computed in their respective
energetic ordering. The Reynolds number is ${\rm
Re}_\tau=125$. } \label{f:mode_energy1}
\end{figure}
One glaring difference between the two flows that stands out from Table 1
is the energy concentration in the few most dominant modes of the
viscoelastic flow compared to the flat distribution of energy in the
Newtonian flow. For example, the first, second and third dominant viscoelastic modes
have all twice the energy of the corresponding first, second
and third dominant Newtonian modes.  The sum of the first four viscoelastic
energies contain as much as as the first ten Newtonian modes. 
The first 15 viscoelastic modes already contain
more than 50\% of the energy, whereas one needs to collect as many as
80 Newtonian modes to reach the same fraction of the total energy.
In Fig. \ref{f:mode_energy1} we display the sum $\C E(n) = \sum_{j=1}^n E(j)$
as a function of $n$ for both flows, where the sum is computed
in the respective energy ordering. We note that the difference between
the two curves is established within the first ten modes; for larger values
of $n$ the lines are almost parallel, indicating a similar energy distribution 
between the less dominant modes. 
\subsection{The energy spectrum}

An interesting and illuminating way to discuss the difference between the
two flows is provided by the energy spectrum. It was proposed in \cite{90Ge}
that the main difference between the spectra of the two flows should
appear in the position of the dissipative scale that separates a spectral
power-law from exponential decay. The increase of the dissipative scale was
indeed observed in recent DNS of homogoneous isotropic turbulence 
in the FENE-P model \cite{02ACBP}. However, the present results indicate that
important changes involve the energy containing scales.

\begin{figure}
\includegraphics[width=8.5cm]{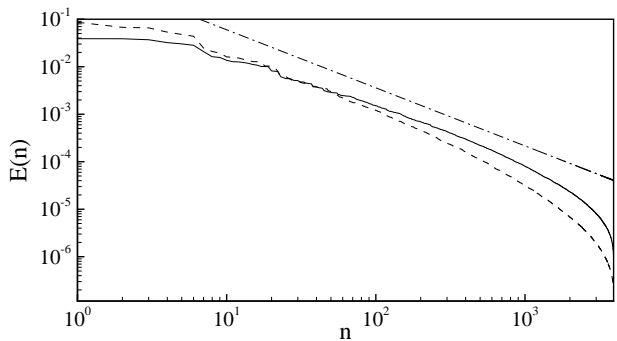}
\caption{Log-log plots of the energy distribution $E_n$ vs mode
index $n$ for the Newtonian case (solid line) and the viscoelastic
case (dashed line). The rough K41 prediction (\ref{K41}) is shown by the
dashed-dotted line. The Newtonian  and visco-elastic dependencies are
shown in their own energetic ordering. The Reynolds number is
${\rm Re}_\tau=125$. } \label{f:mode_energy3}
\end{figure}

In Fig. \ref{f:mode_energy3} the
energy is plotted for the two flows as a function of $n$. The
Newtonian case displays a
spectral plateau for the most dominant modes which crosses
over to a power law for $n_{_{\rm N}}\ge 8$. In the case of the
viscoelastic flow the plateau is dramatically higher, {\em but also the
power law changes its slope}. It appears that the whole spectral curve
is tilted in favor of the energy containing modes and on the expense of the
lower energy modes and the dissipative modes. 

The difference between the power laws can be made clearer by comparing with what
can be obtained from the Kolmogorov law for high $k$-vectors. For relatively large
$k$ we can expect the flow to isotropize \cite{99ALP,02LPT}, and Fourier modes would again
become ``best". In this 
asymptotic situation (neglecting intermittency corrections) the spectrum 
$E(k)$ is expected to be $E(k)\propto k^{-11/3}$. 
In an isotropic environment we also expect that $n$ is proportional
to the volume of the sphere of radius $k$, i.e. $n\propto k^3$.
Thus we can expect for large values of $n$
\begin{equation}
E_n \propto n^{-11/9} \ , \quad n \text{~large}\ . \label{K41}
\end{equation}
The law proposed in Eq. (\ref{K41})
is displayed as the dashed-dotted line in Fig. \ref{f:mode_energy3}.
We see that it is in rough agreement with the power law section of the
Newtonian spectrum, but it is certainly not in agreement with the
viscoelastic spectrum which, as said before, is becoming steeper on the whole. 
This is a clear demonstration of the increase in energy of the energy
containing modes on the expense of the others. We propose that even
the power law section of the viscoelastic spectrum will not have
a ``universal" slope, but rather a slope that depends on the
concentration of the polymer and the degree of drag reduction.

What emerges from this study is that understanding drag
reduction lies in the reordering and energy redistribution of the 
energy containing modes. To examine these phenomena further we consider now
the energy contents of the viscoelastic modes ordered by $n_{_{\rm N}}$ instead
of their own ordering. This plot is a very vivid graphic representation of the 
data of Table 1, stressing very strongly the concentration of energy in the
dominant viscoelastic modes, which are however rearranged in dominance
compared to the Newtonian case.
\begin{figure}
\includegraphics[width=8.5cm]{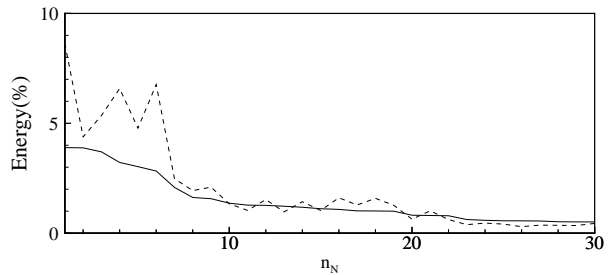}
\caption{Plot of the energy content (in percentage of the total energy)
for the Newtonian (solid line) and viscoelastic (dashed line) empirical modes. 
Both cases are plotted as a function of $n_{_{\rm N}}$.  }
\end{figure}
\begin{figure} 
\includegraphics[width=8.5cm]{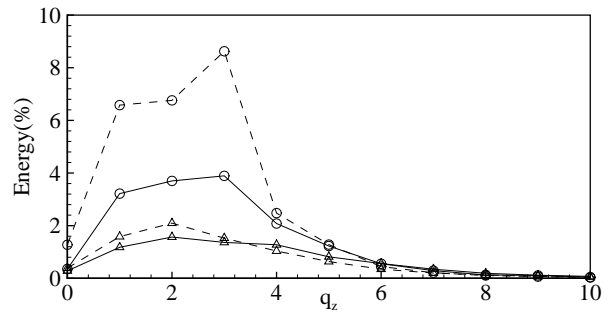}
\caption{Energy of the most energetic modes for a given $\tilde q_x=0$
as a function of $\tilde q_z$. ${\rm p}=1$ modes are represented by circles
and ${\rm p}=2$ modes by triangles. They are connected by solid lines for
Newtonian modes, and by dashed line for viscoelastic modes. }
\label{f:mode_energy5}
\end{figure}
Next we want to reiterate that one should focus on the
most energetic modes. Noticing from Table 1 that all the dominant
modes in both flows are associated with $\tilde q_x=0$, we consider
next these modes as a function of $\tilde q_z$ for $p=1, 2$. In Fig. 
\ref{f:mode_energy5} we show the energy of these modes. We note that
the dramatic redistribution of energy occurs only for the 
most energetic modes with $p=1$; The less energetic modes
are not affected much. Already the modes with $p=2$
are seen in the figure to be affected in a negligible way.
In our opinion this is a clear message that to understand drag
reduction we need to understand the rearrangement of the energy
containing modes. For our flow configuration the most relevant are the modes
which are space homogeneous in the spanwise direction. It should be stressed
however that different geometries, and even channel geometry with different
aspect ratios, may bring forth other modes as the most relevant ones.
Nevertheless we expect that drag reduction would always be associated
with a substantial increase in the energy containing modes, whatever
these are for a given flow configuration.
\section{conclusions}
\label{summary}
In this paper we initiated a systematic study of drag reduction on
the basis of DNS of the FENE-P model. We investigated simulations of
Newtonian and viscoelastic flows in channel configuration at
the same friction Reynolds number. Our main aim is to understand
the mechanism of drag reduction. Since drag reduction involved
modifications of the mean flow and of the large scale gradients
in the flow, we are motivated to understand more the energy containing
modes, rather than focusing only on phenomena of small scales. To this
aim we have found first the list of empirical modes that represent the
velocity field in an optimal way in an energy decreasing ordering.
The first important discovery was that this list contains the very
same modes for the Newtonian and viscoelastic flows. We propose that
this finding will offer a huge simplification in any future
theory of drag reduction. While we cannot offer a clean explanation
of this finding, we can proceed at this time taking the approximate
invariance of the modes as an empirical fact. What needs to be understood is just the
energy distribution and reordering of the modes in the viscoelastic
case. 

We should stress that the point of view proposed here differs
in a fundamental way from the approach presented for example in Refs.
\cite{69Lu,90Ge,01BFL}. The thinking there focuses on the energy cascade in
the turbulent flow, and on the modification of the small dissipative scales.
Roughly speaking, the maximal gradients of the velocity are estimated
from balancing the RHS of Eq. (\ref{EqR}). Neglecting the statistical correlation
between the conformation tensor and the velocity gradient, one can estimate
the maximal velocity gradient by $\langle \partial u/\partial r\rangle \sim 1/\tau_{\rm p}$.
Thus the matching of the polymer relaxation time with an eddy turn over time
is used to predict a decrease in the maximal velocity gradient which is
interpreted as drag reduction. We take an exception to this approach. First,
a careful analysis of the space dependence of the conformation tensor
shows that it is highly correlated with the velocity gradients (see for
example \cite{98DSB,02ACBP,02BP}. Thus the estimate taken above
is questionable at best. But moreover, we have shown that the main changes
between the Newtonian and the viscoelastic modes occur in the energy
containing modes. 
The energy containing modes are highly anisotropic, they are not Fourier
modes, and their connection to the small scales where the isotropised Kolmogorov picture
is tenable is very unclear. A theory that assumes a K41 spectrum down to
a modified viscous scale does not appear tenable for the FENE-P flow,
as seen in Fig. \ref{f:mode_energy3}.  We propose that a theory of drag reduction
entails an understanding of the relative energy of the modes that
characterize the largest scales of the flow.

To understand the relative ordering and the relative energy of the most
dominant modes one needs to study the energy intake by these modes 
from the mean flow \cite{01GLP,02GLP}, the energy exchange between the modes, and the
energy exchange with the viscoelastic subsystem represented by the
conformation tensor field $\B R(\B r,t)$ \cite{02ACBP,02BP}. Such an investigation calls
for measuring additional statistical objects like 3rd order correlation
functions. The necessary simulation are in progress and the results
will be published elsewhere. 

\acknowledgements 
We are indebted to Anna Pomyalov for 
helpful suggestions in the course of this work. This work has been
supported in part by the European Commission under a TMR
grant, the German Israeli Foundation, the Minerva Foundation, 
Munich, Germany, the Israeli Science Foundation and the
Naftali and Anna Backenroth-Bronicki Fund for Research in
Chaos and Complexity.

\end{document}